# Water under the Ridge: Evaporation, Translation, Crumpling and Encapsulation of a Water Droplet atop a Liquid Polymeric Film

*Sri Ganesh Subramanian\*, Sachin Nair\*, Sunando DasGupta[1]*

*Department of Chemical Engineering, Indian Institute of Technology Kharagpur, 721302, India*

[1]Corresponding author

\*Equal Contribution

Email: sunando@che.iitkgp.ernet.in

Ph.: +91 - 3222 - 283922

# ABSTRACT:

The intriguing dynamics of a compound liquid-system comprising of water and (uncured) poly-dimethylsiloxane is explored in the present work. The viscoelastic nature of the film, coupled with the dynamics of evaporation, triggered a self-propulsion in the droplet, which gradually segued into the crumpling of the film, and finally culminated in the encapsulation of the water droplet by the polymer. The physics of the hitherto unreported phenomena has been explained via the development of an analytical model, by taking into account all the germane forces. It is conjectured that this symbiotic and self-sustained dynamics, aided with the non-requirement of any complex fabrication procedures, would pave the path for the development of precision drug-delivery, unmediated flow-focusing, self-mixed micro-reactors, the study of micro-swimmers, surface encapsulation, and photonics, to name a few.

# INTRODUCTION:

The inception of the investigation of evaporation could be dated back to early 18$^{th}$ century, and subsequently, numerous researchers have contributed substantially towards understanding the process of evaporation in general and droplet evaporation in particular.[1–4] However, the physics underlying the evaporation of a droplet atop a *viscoelastic liquid film* still remain nebulous. The cognizance of the dynamics pertaining to droplet evaporation is imperative to several applications such as heat transfer,[5] spray-painting,[6] ink-jet printing[7] etc. More specifically, for tailored applications involving Lab-on-a-chip (LoC) devices and Point-of-Care (PoC) diagnostic systems, the dynamics of evaporation needs to be well comprehended, in order to circumvent the subtleties that would arise during the handling of open-systems.[8] Moreover, the effect of substrate and liquid properties would play a crucial role in ascertaining the dynamics of such systems. However, there appears to be a dearth of literature pertaining to the evaporation of droplets in contact with another liquid, as the substrate,[9–11] as opposed to evaporation on rigid or soft-substrates.

The present work (which is a *first-of-its-kind* study, to the best of the knowledge of the authors) includes a wide array of intriguing dynamics that focuses on the properties of the liquid and the viscoelasticity of the polymer (liquid film, which acts as the substrate). Unlike a SLIP (slippery liquid-infused porous) surface, wherein photolithography or any other soft lithographic method has to be employed in order to hold the viscous liquid,[9,12,13] the present system, owing to the inherent nature of the viscoelastic liquid, has no such requirement of additional surface modification or fabrication, and hence the surface could be termed as – *slippery viscoelastic liquid surface (SVELS)*. Furthermore, the analysis of the pertinent dynamics is broken-down into two regimes: one which involves an evaporation phase, followed by the second phase which is free from evaporation and is completely focused on wrinkling and crumpling of the polymeric film. The investigation of such multi-functional systems would provide engineering solutions towards the development of Point of Care diagnostic devices. The need for the development of such a multi-functional PoC device, coupled with the scientific inquisitiveness, forms the motivation for the present work.

The dynamics of evaporation was studied with DI water (*Millipore India Pvt. Ltd*) as the solvent and un-cross-linked poly-dimethylsiloxane (PDMS) (*Part-A of Sylgard -184, Dow Corning, USA*) as the polymeric film, with glass as the base substrate (***see Supplementary Information SS2***). The entire sets of experiments were carried out on a laser scanning confocal microscope (LSCM) (*LEICA, DMI 6000 CS)*. Contact angle and base-radius measurements of the droplet were undertaken using a goniometer (*Data Physics, OCA 15 Pro*). The thickness of the polymeric film was fixed to be $13 \pm 0.01$ μm and three different volumes (0.3, 1, and 2 μl) of the droplet were used for the present work. Extreme caution was exercised towards placing the droplets on the polymeric film (by using an automatic dispensing system, and using a pre-calibrated syringe (*Hamilton*)); to ensure symmetrical deposition, and asymmetrically placed droplets were disregarded from the analysis (post experimentation). Quintuple readings were taken for each volume-thickness combination to ensure repeatability of the observed phenomena. The dynamics were recorded throughout the entire length of the experimentation time, and the obtained videos were converted into images with a time-span of 1.1s and were analyzed using MATLAB (v. R 2015 b) subroutines. The entire dynamics of the pertinent phenomena is elucidated sequentially, in the order of their inception.

# RESULTS AND DISCUSSION:

The evaporation of water, atop the polymeric film, leads to a temporal variation in the radius of the droplet, giving rise to a macroscopic translation, and propulsion of the droplet. The process gradually segues into the crumpling of the polymeric film, and finally culminates after the encapsulation of a minute quantity of the remnant liquid by the polymeric film, as depicted in Figure 1. The present section has been broadly classified into two sub-sections: the first sub-section elucidates the dynamics of evaporation and the macroscopic movement of the droplet; whereas, the second sub-section entirely focuses on the dynamics of crumpling and encapsulation, thereby providing a sequential outline of the apposite physics.

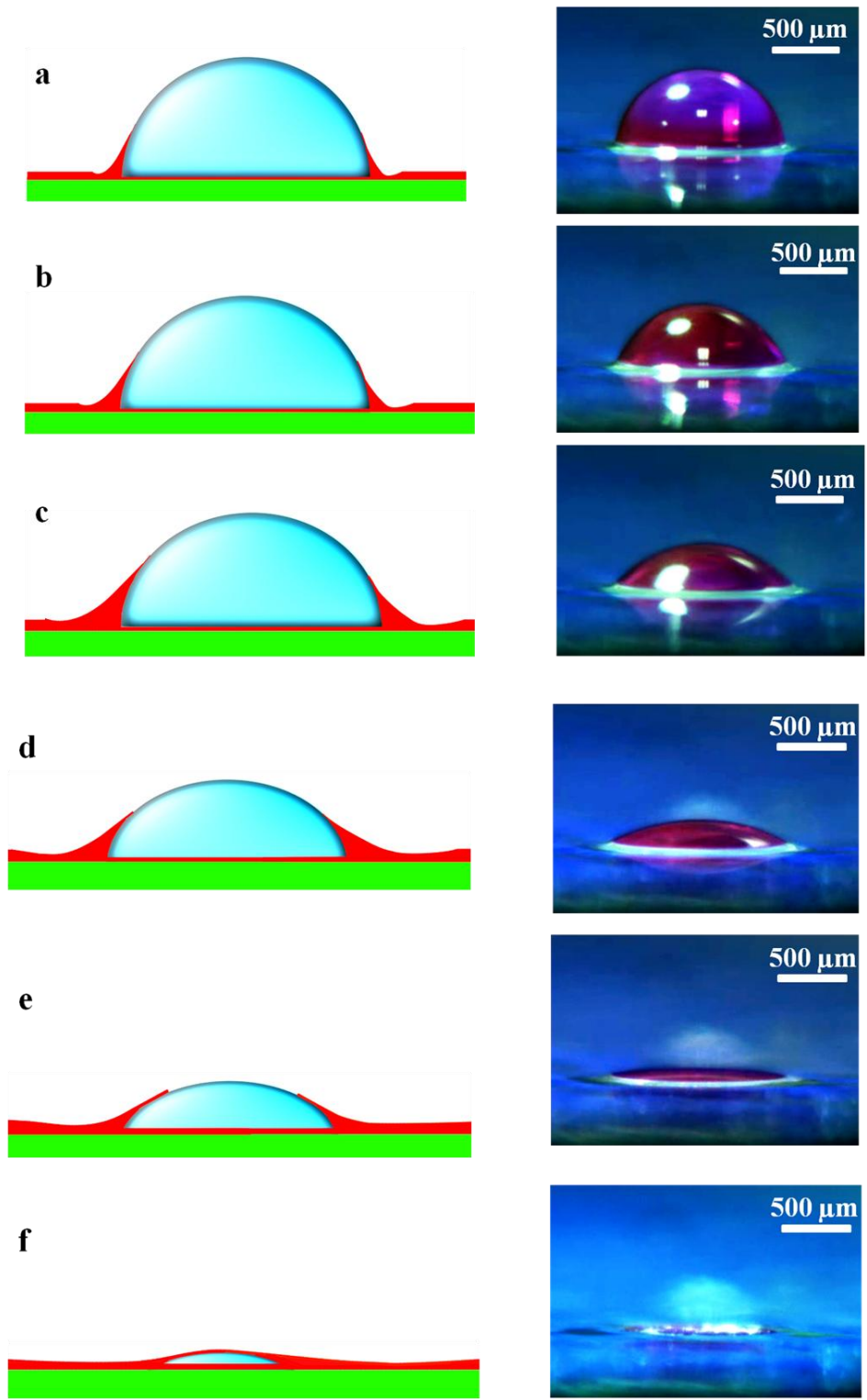

Figure 1: Schematic (left) and experimental (right) images (volume = 1μl) of the temporal sequence, involving evaporation and encapsulation (*encapsulation is experimentally not perceptible here, refer Supplementary Information SS1*). Notice how the bright ridge at the periphery of the droplet slowly increasing in width and flows over the droplet from the sides,

especially in the sequence d-e-f. In order to discern the droplet from the film, Rhodamine-B was added to the droplet.

**EVAPORATION AND TRANSLATION**

The radius of the droplet has been used to determine the rate and dynamics of evaporation. The variation in the non-dimensional radius and contact angle of the droplet, as a function of time, is depicted in Figure **S 2.1 and S 2.2** (*see Supplementary Information SS3*). The radius and the contact angle were normalized by considering their initial values at t = 0s, as the reference. Figure (2.1) depicts the dynamics of the droplets when the radiuses were plotted as a function of logarithmic variation in time, thereby yielding an interesting orientation. It could be inferred from the graph that, all the data, irrespective of the volume, follow a similar trend, which is characterized by the existence of two well-defined regimes: In the first regime, spanning $\sim 10-15 s$, it could be observed that there is a near-constant variation in the radius of the droplet. However, as the time progresses, the dynamics transform into a falling rate regime, wherein there is a rapid reduction in the radius of the droplet. The initial constancy in the radius could be attributed to the time taken by the droplet to spread over the film and attain an equilibrium state. However, the film being viscoelastic in nature, with a characteristic-time in the order of few milliseconds ($3.67 \pm 0.05$ ms), would induce a local time lag, resulting in a dynamic, yet slow equilibration of the droplet.[14–18] Once the droplet attains its initial equilibrium, the radius (of the droplet) follows a logarithmic decrease. The present dynamics is significantly different from the constant-contact-angle (CCA) and constant-contact-radius (CCR), regimes of droplet-evaporation,[2,4,9,19] in the sense, that there is a faster decrease in the radius of the droplet, in comparison to the dynamics over a rigid or a soft substrate. However, this does not translate to a decrease in the evaporation-time because; the liquid-vapor surface area of the droplet gets progressively covered by the polymeric film, thereby reducing the surface area available for evaporation. Hence, the overall time taken for evaporation of the droplet in the present case is comparable to that over a soft substrate (*see Supplementary Information SS 6*).

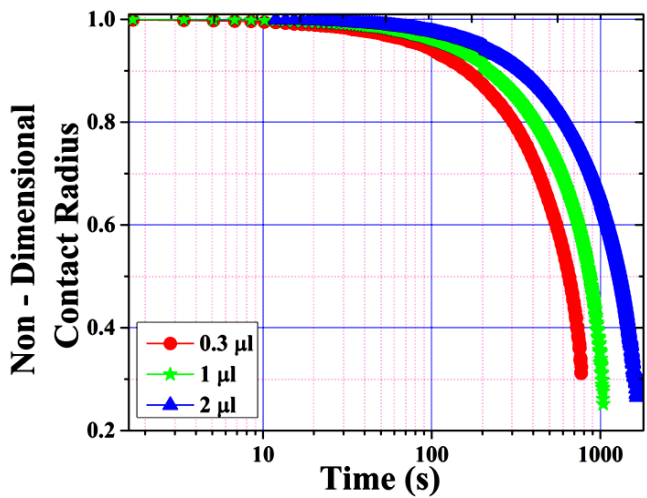
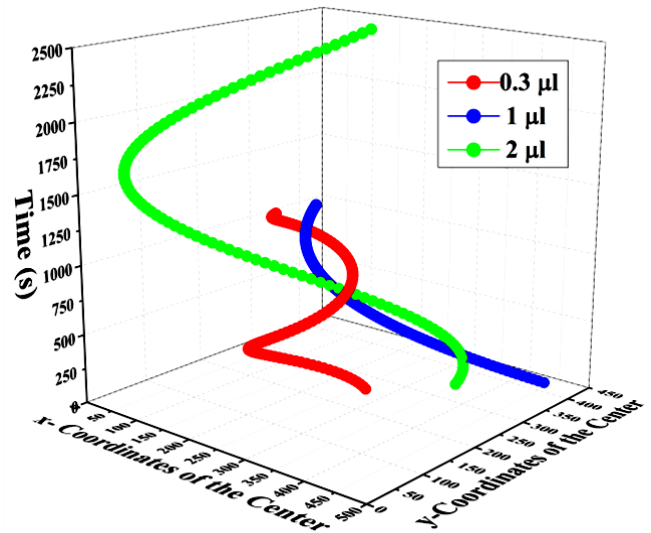

(2.1)                                  (2.2)

Figure (2.1): Non-dimensional variation in the Contact-radius as a function of logarithmic time-scale, clearly discerning the different regions of evaporation. Figure (2.2): Temporal variation in the coordinates of the center of the droplet indicating the self-propulsion during evaporation on a slippery viscoelastic liquid (SVEL) film.

Figure 2.2 denotes the movement of the droplet, as a function of time (represented in the z-axis). It is to be noted that this representation is only to enhance the readability of the graph, and as such, the droplet follows a non-linear (curved) path and not a helical one (*see Supplementary Information SS4*). The actual translation of the droplet is depicted in Figure 3.1 (*see Supplementary Information SS1.a and SS1.b*), and the corresponding coordinate positions are depicted in Figure **S 3.1**. It is interesting to observe that; the droplets continue to follow a non-linear path (resembling a curve). However, the droplets tend to reverse their direction, after a particular amount of time and this could be conjectured to a change in the circulative pattern (flow pattern) inside the droplet, caused due to a steady decrease in the available surface area.[11]

Furthermore, the critical reason which could be attributed to the non-linear motion of the droplet is the varying curvatures, across the free-surface of the droplet (across the PDMS-droplet and droplet-air interfaces). Although the droplet is macroscopically symmetric, the effect of viscoelasticity of the polymer renders the droplet to be in a state of microscopic asymmetry, such that there is an observable difference in the left and right-contact angles, measured at every differential element of time, along the three-phase contact line (TPCL). At any given instance of time, a portion of the contact line retracts faster as compared to the other half, as depicted in Figure **S 3.2**. For the case of solid substrates or for soft-surfaces in general, this small difference has no direct consequence on the movement of the droplet, due to the inherent effect of pinning, or other resistive forces (*see Supplementary Information SS1.c*).[20–22] On the other hand, this difference could become a considerable impetus if the magnitudes of the resistive forces are lower in comparison to the force generated due to the asymmetry of the droplet. More importantly, during the retraction phase, it could be observed that a small region of the TPCL would recede at a slower velocity than the other regions, as evident from the fluctuations in the velocity of the droplet, as depicted in Figure **S 3.2**. This relative velocity between successive intervals of time creates a dynamic difference in the capillary pressure, causing the droplet to translate. In other words, consider a spherical rubber ball over a handkerchief, which is held across two ends. If a change in curvature of the handkerchief is introduced across one of its ends, it could be observed that the ball moves to the other end, in response to the induced change. Similarly, a droplet atop a SVEL surface would try to move, even if there is the slightest asymmetry in the curvature, due to the deformation across the TPCL.

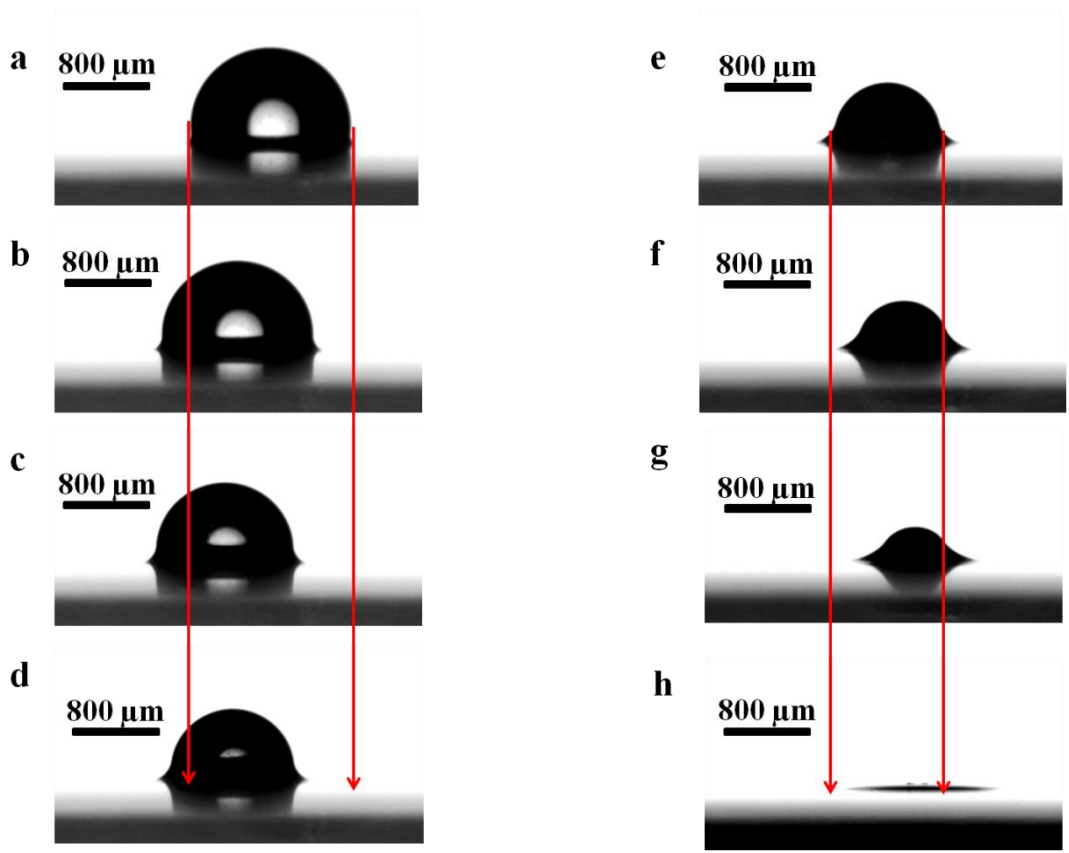

(3.1)

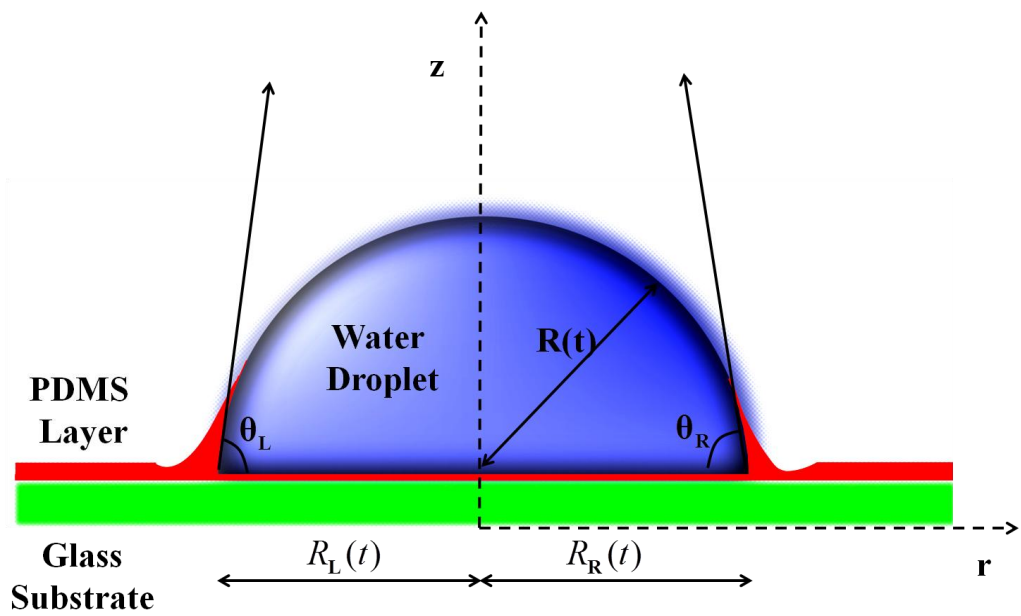

(3.2)

Figure (3.1): Optical side-view projection of the self-propulsion of the droplet. The images clearly illustrate a reversal in direction of the droplet with time. Another clear feature of the side view is the apparent increase in height of the wetting ridge as the height of the spherical droplet progressively decreases. Figure (3.2): Schematic for theoretical modeling of the droplet dynamics.

In order to evince the existence of such a universal impetus, all the pertinent forces are taken into consideration. It is to be noted that, for the system under consideration, the capillary length $\kappa_{ij} = \sqrt{\frac{\gamma_{ij}}{\Delta \rho_{ij} g}} \geq 0.05 m$ (where, (i, j, k), denote the cyclic combination of the three phases involved), whereas the maximum length-scale (for a droplet volume = 2 µl) for the droplets is around L = 2.25mm, resulting in the corresponding Bond numbers $B_{ij} \cong (L/\lambda_{ij})^2 \approx 0.001 \ll 1$. This implies that the effect of gravity could be safely omitted while considering the equilibrium shape of the droplets. The force inducing the droplet motion arises due to an inherent variation in the curvature across the two halves of the droplet, as depicted in Figure 3.2. Albeit the process is dynamic, the variation in curvature is significantly slow, and hence a quasi-steady state model to underline the inherent physics has been considered. The resistive forces include contributions from the pinning force,[20,22,23] the drag force due to the ambiance,[21,24,25] the viscous dissipation of the droplet[25,26] and most importantly, the viscoelastic dissipation due to the propagation of the wetting ridge.[25–27] At equilibrium, the net force could be represented as,

$$F_{net} = F_{Driving} - F_{Pinning} - F_{Drag} - F_{LDiss} - F_{VEDiss} \quad \ldots \quad (1)$$

The driving force could be resolved by considering an inherent pressure difference between the two halves (the left and the right sides of the z-axis in Figure 3.2) of the droplet. This pressure difference arises due to a dynamics difference between the advancing and receding contact angles of the droplet. The magnitude of this force could be calculated as,

$$F_{Driving} = A_{slL}\frac{\gamma_{lv}}{R_L} - A_{slR}\frac{\gamma_{lv}}{R_R} \quad \ldots (2)$$

where, $A_{slL}$ and $A_{slR}$ are the solid-liquid interfacial areas of the left and right halves of the droplet, $R_L$ and $R_R$ are the radii of the left and right side of the z-axis depicted in Figure 3.2 respectively, and $\gamma_{lv}$ is the liquid-vapor interfacial tension.

The procedure employed for calculating the values of the radiuses, and the governing equations for the resistive forces are presented in ***Supplementary Information SS 5***. Therefore, the net force could be represented as,

$$\begin{aligned}F_{Net} =& \left(A_{slL}\frac{\gamma_{lv}}{R_L} - A_{slR}\frac{\gamma_{lv}}{R_R}\right) - \left(\pi R_M \gamma_{lv}\sin\theta[\cos\theta_L - \cos\theta_R]\right) - \frac{1}{2}C_D\rho_f A_{sl}U^2 - \frac{3\eta U^2 ft}{\tan\theta} \\ & - \frac{2\psi\gamma_{lv}^2(1-\upsilon^2)U}{\pi G^*}\left\{\frac{1}{\Phi} + \frac{3\theta^2}{R_M}\left[2\ln\left(\frac{R_{max}}{\Phi}\right) - 1\right]\right\}t \end{aligned} \quad \ldots (3)$$

where, $R_M$ is the contact radius of the droplet (average of left and right radius), $\theta$ is the mean contact angle, $C_D$ is the drag coefficient of the droplet, $\rho_f$ is the density of the ambience, $U$ is the dynamic-velocity of the TPCL, $\eta$ is the viscosity of the droplet, $f$ is a functional relationship between the minimum and maximum length-scales, $t$ denotes the time, $\psi$ denotes the fraction of the total energy expended as viscous dissipation, $\upsilon$ is the Poisson's ratio of the polymer, $G^*$ is the complex Young's modulus of the polymer, $R_{max}$ is the radius of the droplet at t=0, and $\Phi$ is the minimal distance from the edge of TPCL wherein non-linear effects predominate.

The temporal variation of the apposite forces is presented in Figure 4.1. It is evident from the plot, that the driving force follows a curved path, and is not linear. Therefore, it would be cogent to infer that the reason for the non-linear translation of the droplet is due to the meandering nature of the driving force, arising due to a variation in the curvature, across the two ends of the drop. Furthermore, it could be deduced that the major resistive forces are the pinning and the viscoelastic dissipation and as the droplet continues to evaporate, the film continues to cover a greater portion of the liquid-air interface, and hence the contribution due to the viscoelastic dissipation continues to increase, preventing the drop from further movement. In order to verify the present postulation and also to extend the prediction to any general viscoelastic liquid, the experiments were repeated on a variety of substrates such as silicone oil (purely viscous), soft PDMS (purely elastic), and Teflon (viscoelastic) (*see Supplementary Information SS1.c, SS1.d, and SS1.e*). The dynamics of the droplet and the corresponding net force plots are presented in the *Supplementary Information SS6*. The droplet dynamics on these substrates is a clear indication of the existence of a universal impetus, which manifests itself in the absence of any dominant resistive force.

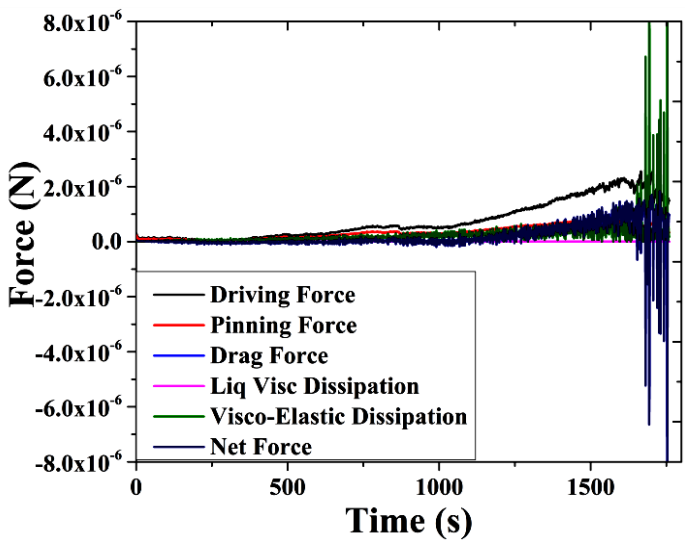

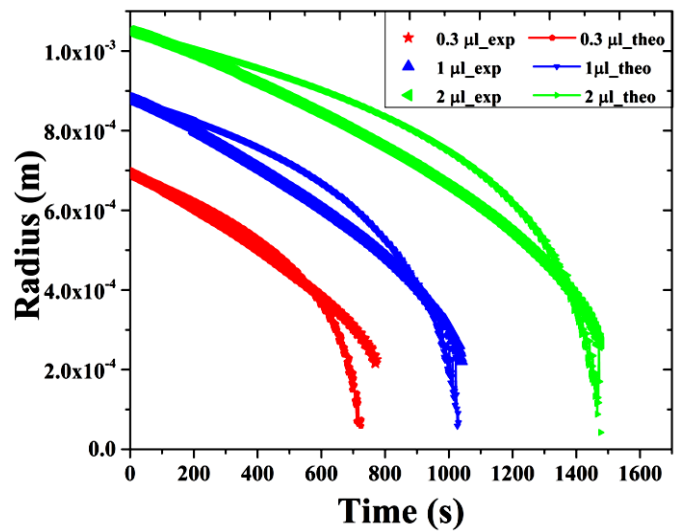

(4.1)                                                                 (4.2)

Figure (4.1): Temporal variation in the forces affecting the dynamics of the droplet. The net driving force is non-linear, explaining the curved path of the droplet. Figure (4.2): Comparison between theoretical and experimental radii of the droplet.

Since the droplet evaporates, the dynamic variation in the radius of the droplet could be modeled using a diffusion-based model, in the absence of any external variation in the ambient conditions (*see Supplementary Information SS 5*). The experimental results are indicative of the fact that the droplet adopts a spherical shape, throughout the length of the experimentation time. Therefore, the volume of the droplet could be calculated by considering the droplet to be a part of a spherical cap, which would result in a governing equation of the form[2,9,28]:

$$\frac{dV}{dt} = -\frac{D}{\rho}\int(\nabla \cdot c)dA_{lv} \quad \quad \ldots \quad (4)$$

where, V is the volume of the droplet, D is the coefficient of diffusion and $(\nabla \cdot c)$ is the gradient in the concentration of the vapor along the liquid-vapor interface.

Although the contact angle varies dynamically, the inclusion of such a variation poses inherent complications in the determination of a theoretical radius, and hence, for brevity, the variation in the dynamic contact angle is assumed to be less dominant, in relation to that of the contact radius.[9]

The final equation depicting the temporal variation in the radius of the droplet is given as,

$$R(t) = \left(R^2_{max} - \left(\left\{\frac{6D(C_s - C_\infty)}{\rho}\right\}\left\{\frac{p(\theta)(1+\cos\theta)^2 t}{\sin\theta(2+\cos\theta)}\right\}\right)\right)^{1/2} \quad \quad \ldots \quad (5)$$

The comparison between the experimental and the theoretically predicted radius of the droplet is depicted in Figure (4.2). Although the present model provides a reasonable agreement with the experimental results (error: 3 - 9 %), the overprediction could be attributed to the non-inclusion of a dynamic variation in the contact angle, and the translative motion induced by the internal pressure,[11] which would be a part of the future work. The droplet, after propelling for a considerable distance, would now gradually transition towards crumpling of the film, as elucidated subsequently.

## CRUMPLING AND ENCAPSULATION:

The dynamics of droplet evaporation, as described previously, proceeds to a final stage, resulting in the formation of wrinkles at the periphery of the TPCL, leading towards crumpling of the polymeric film, as depicted in Figure 5. Although the previous studies include the wrinkling of polymeric films placed over a spherical droplet,[29–33] the present study on the other hand, has a different configuration wherein, the droplet rests on top of a polymeric film, and during evaporation, the polymeric film wrinkles and crumples, thereby encapsulating a portion of the liquid volume within itself (*see Supplementary Information SS 1.i*).

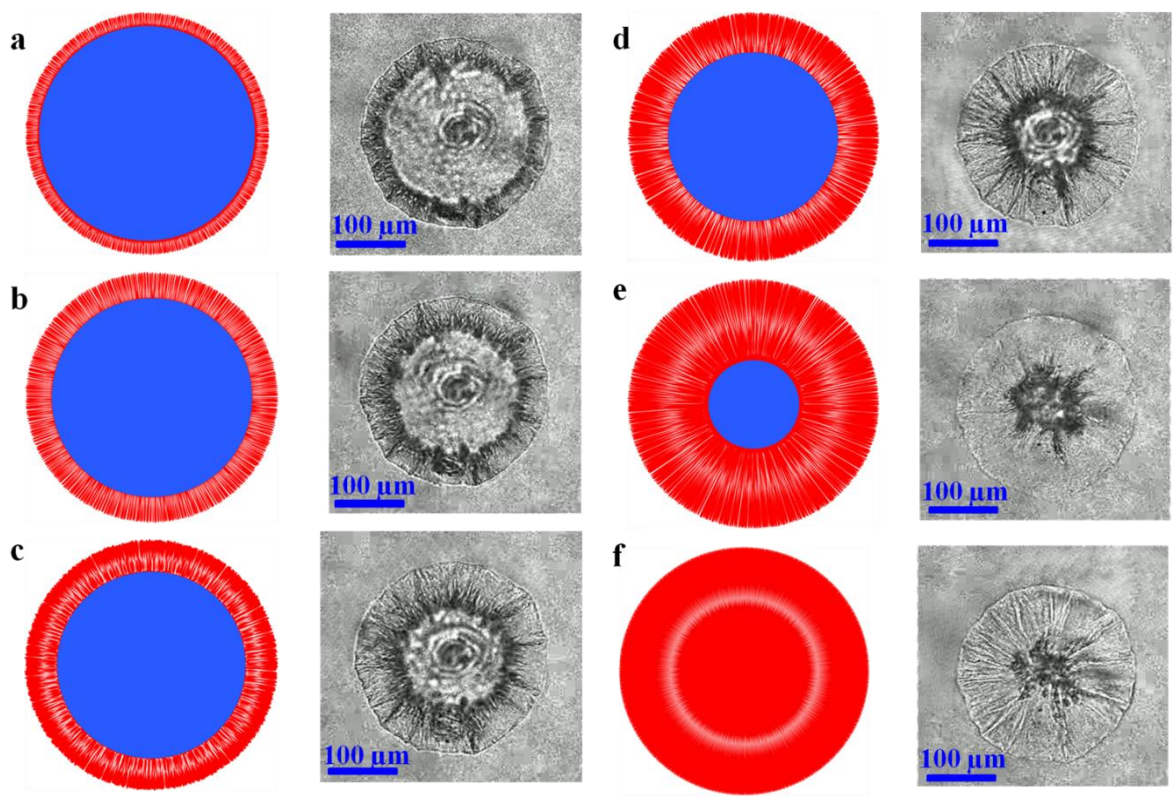

Figure 5: Schematic and Laser Scanning Confocal Microscope imagery of the crumpling and encapsulation phenomena.

The LSCM images and the high-speed camera images of droplet evaporation (of water containing a fluorescent dye: Rhodamine-B) (*see Supplementary Information SS 1.j, SS 1.k and SS 1.l*) optically confirm the encapsulation by the polymeric film, in the later stages of evaporation. However, in order to ascertain the existence of water inside the polymeric film post-encapsulation, fluorescent spectroscopy and ATR-FTIR were used (*see Supplementary Information SS 7*). Furthermore, in order to assess the effect of the substrate thickness on the dynamics of encapsulation, the thickness of the PDMS film, was increased to 90 μm (*see Supplementary Information SS 1.m*). In addition, as a separate study, droplets containing neutrally buoyant fluorescent particles, at a very low concentration (0.001 % (w/v)) (polystyrene, 1 um diameter, (Sigma)), were allowed to evaporate atop the SVEL surface, to dynamically view the encapsulation process (*see Supplementary Information SS 1.n*). These experiments qualitatively confirm the existence of water inside the polymeric film, the quantitative determination, however, is beyond the scope of the present work and would be undertaken as a part of the future study. Moreover, in order to verify the energetic of the system, numerical simulation using Surface Evolver[34] were undertaken (*see Supplementary Information SS 8*). Although the encapsulation of water by PDMS is not energetically favorable, the data obtained from the simulations indicate that the system approaches a local minima and not a global minimum in terms of energy, which suggests that the complex interplay between the dynamic wetting ridge, and the decreasing volume of the droplet, inadvertently causes the film to encapsulate a minute quantity of water.

Although there appears to be, only a handful of literature that addresses the theoretical formulation and physics involved,[29,30] a generic route wherein a polymeric-film crumples and thereby encapsulates another liquid, still remains elusive. It is overt that a viscoelastic liquid can release an applied stress in one of the two possible ways – either by deforming (elasticity) or by flowing (viscosity). In the current scenario, wrinkling is observed in an annulus just outside the diameter of the droplet, towards the end of the evaporation. This region corresponds to the descent of the wetting ridge formed just outside of the TPCL, as depicted in Figure 5. The TPCL constantly exerts a tension on the free PDMS layer (due to the normal component of the surface tension), to which the free PDMS layer responds elastically, creating a ridge. The TPCL also moves inwards with a varying velocity, which implies that at every differential element of time, a new ridge is formed. The movement of the TPCL causes two things: (a) It pulls onto the PDMS layer just ahead of it, and (b) stretches the layer just behind it, thereby propelling the droplet forward.[16,35] During the later stages of the evaporation, the velocity of the contact line gradually increases and attains a significantly higher value, sufficient enough to pull the wetting ridge so quickly that the elasticity causes an azimuthal deformation of the decent behind the ridge, resulting in the formation of wrinkles. The compression is relieved by an initial transformation into a wrinkle pattern, and then into a crumpled state via a continuous transition. The present work supports the hypothesis of King et al,[29] which states that the wrinkle formation is not a final asymmetric configuration, but is an energetically favored interim state, post which the film would begin to crumple. The surrounding PDMS would then, by virtue of its viscosity, flow inwards to relax the excess strain energy, causing the wrinkles to vanish, ultimately resulting in the crumpling of the polymeric film, thereby encapsulating a tiny amount of liquid (water) within. Although these observations include contribution from the asymmetry of the droplet, and the associated stress-fields, a generic model to address the inherent dynamics, warrants a detailed study on the crumpling and encapsulation alone, taking into account, the nature of the material, the thickness of the film, and the volume of the droplet, which would be a part of the future work.

## CONCLUSION:

The dynamics of an evaporating droplet atop a liquid viscoelastic polymer was investigated. The evaporation follows a dynamic decrease in both, contact radius and contact angle, as opposed to the existing CCR and CCA modes of droplet evaporation. The present study forms the basis for a new domain of inquiry, with self-propulsion, wrinkling, crumpling and encapsulation, all occurring within the same system, sequentially. These dynamics are a result of the self-sustained and symbiotic relationship between the evaporation of the droplet and the viscoelastic property of the polymeric film. Furthermore, unlike a SLIP surface, the present configuration, owing to the properties of the liquid film, does not necessitate the need for complex fabrication procedures, thereby giving rise to a SVEL surface. A theoretical model was developed to explain the dynamics of the translating droplet, and the associated temporal variation in the radius. It is postulated that the present study would find significant applications in several novel and existing microfluidic systems, owing to its potential and inherent multi-functionality.

**ACKNOWLEDGEMENTS**

One of the authors SSG, would like to take this opportunity to extend his heartfelt gratitude to Prof. Dibakar Dhara and Ms. Puja Poddar, Department of Chemistry, IIT Kharagpur, for their suggestions and support in characterization of the samples; Mr. Banuprasad T. N, Advanced Technology Development Centre, IIT Kharagpur, for the suggestions and help with the high-speed camera imagery, and for all the useful discussions; The members of the MTP Lab, Department of Chemical Engineering, for their suggestions; Mr. Druba Shakha, DRF, Department of Chemical Engineering, IIT Kharagpur, for his aid in FTIR studies; Mr. Siddhartha Mukherjee, Advanced Technology Development Centre, IIT Kharagpur, for the measurements with the rheometer.

**REFERENCES**

1. Picknett, R. G. & Bexon, R. The Evaporation of Sessile or Pendant Drops in Still Air. *J. Colloid Interface Sci.* **61,** 336–350 (1977).

2. Stauber, J. M., Wilson, S. K., Duffy, B. R. & Sefiane, K. On the lifetimes of evaporating droplets. *J. Fluid Mech.* **744,** 1–12 (2014).

3. Elbaum, M. & Lipson, S. G. How does a thin wetted film dry up? *Phys. Rev. Lett.* (1994). doi:10.1103/PhysRevLett.72.3562

4. Brutin, D. *Droplet Wetting and Evaporation. From Pure to Complex Fluids*. (Elsevier,

2015).

5.  Paik, P. Y., Pamula, V. K. & Chakrabarty, K. Adaptive Cooling of Integrated Circuits Using Digital Microfluidics. *IEEE Trans. VERY LARGE SCALE Integr. Syst.* **16,** 432–443 (2008).

6.  Sirignano, W. A. *Fluid Dynamics and Transport of Droplets and Sprays*. (Cambridge University Press, 2010).

7.  Fobel, R., Kirby, A. E., Ng, A. H. C., Farnood, R. R. & Wheeler, A. R. Paper microfluidics goes digital. *Adv. Mater.* **26,** 2838–2843 (2014).

8.  Long, Z., Shetty, A. M., Solomon, M. J. & Larson, R. G. Fundamentals of magnet-actuated droplet manipulation on an open hydrophobic surface. *Lab Chip* **9,** 1567–1575 (2009).

9.  Guan, J. H. *et al.* Evaporation of Sessile Droplets on Slippery Liquid-Infused Porous Surfaces ( SLIPS ). *Langmuir* **31,** 11781–11789 (2015).

10. Sun, W. & Yang, F. Evaporation of a Volatile Liquid Lens on the Surface of an Immiscible Liquid. *Langmuir* **32,** 6058–6067 (2016).

11. Gelderblom, H., Stone, H. A. & Snoeijer, J. H. Stokes flow in a drop evaporating from a liquid subphase. *Phys. Fluids* **25,** 102102 (1-15) (2013).

12. Hui Guan, J. *et al.* Drop transport and positioning on lubricant-impregnated surfaces. *Soft Matter* **13,** 3404–3410 (2017).

13. Wong, T. S. *et al.* Bioinspired self-repairing slippery surfaces with pressure-stable omniphobicity. *Nature* **477,** 443–447 (2011).

14. Kajiya, T. *et al.* Advancing liquid contact line on visco-elastic gel substrates: stick-slip

vs. continuous motions. *Soft Matter* **9,** 454–461 (2013).

15. Style, R. W., Hyland, C., Boltyanskiy, R., Wettlaufer, J. S. & Dufresne, E. R. Surface tension and contact with soft elastic solids. *Nat. Commun.* **4,** 1–6 (2013).

16. Karpitschka, S. *et al.* Droplets move over viscoelastic substrates by surfing a ridge. *Nat. Commun.* **6,** 7891(1-7) (2015).

17. Andreotti, B. *et al.* Solid Capillarity: When and How does Surface Tension Deform Soft Solids? *arxiv:1512.08705v1* 1–5 (2015).

18. Marchand, A., Das, S., Snoeijer, J. H. & Andreotti, B. Contact angles on a soft solid: From young's law to neumann's law. *Phys. Rev. Lett.* **109,** 1–5 (2012).

19. Yu, D. I. *et al.* Dynamics of Contact Line Depinning during Droplet Evaporation Based on Thermodynamics. *Langmuir* **31,** 1950–1957 (2015).

20. Luo, J. T. *et al.* Slippery Liquid-Infused Porous Surfaces and Droplet Transportation by Surface Acoustic Waves. *Phys. Rev. Appl.* **7,** 1–9 (2017).

21. Chakraborty, M., Ghosh, U. U., Chakraborty, S. & Dasgupta, S. Thermally enhanced self-propelled droplet motion on gradient surfaces. *RSC Adv.* **5,** 45266–45275 (2015).

22. Banuprasad, T. N. *et al.* Fast Transport of Water Droplets over a Thermo-Switchable Surface Using Rewritable Wettability Gradient. *ACS Appl. Mater. Interfaces* **9,** 28046–28054 (2017).

23. Snoeijer, J. H. & Andreotti, B. Moving Contact Lines : Scales , Regimes , and Dynamical Transitions. *Annu. Rev. Fluid Mech.* **45,** 269–294 (2013).

24. Pimienta, V. & Antoine, C. Self-propulsion on liquid surfaces. *Curr. Opin. Colloid Interface Sci.* **19,** 290–299 (2014).

25. Shanahan, M. E. R. The spreading dynamics of a liquid drop on a viscoelastic solid. *J. Phys. D Appl. Phys* **21,** 981–985 (1988).

26. Shanahan, M. E. R. & Garré·, A. Viscoelastic Dissipation in Wetting and Adhesion Phenomena. *Langmuir* **11,** 1396–1402 (1995).

27. Pandey, A. *et al. Dynamical Theory of the Inverted Cheerios Effect*. (2017). doi:10.1039/C7SM00690J

28. Popov, Y. O. Evaporative deposition patterns: Spatial dimensions of the deposit. *Phys. Rev. E - Stat. Nonlinear, Soft Matter Phys.* **71,** 1–17 (2005).

29. King, H., Schroll, R. D., Davidovitch, B. & Menon, N. Elastic sheet on a liquid drop reveals wrinkling and crumpling as distinct symmetry-breaking instabilities. *Proc. Natl. Acad. Sci.* **109,** 9716–9720 (2012).

30. Vella, D., Adda-bedia, M. & Cerda, E. Capillary wrinkling of elastic membranes. *Soft Matter* **6,** 5778–5782 (2010).

31. Huang, J. *et al.* Capillary Wrinkling of Floating Thin Polymer Films. *Science (80-. ).* **317,** 650–653 (2007).

32. Cerda, E. & Mahadevan, L. Geometry and Physics of Wrinkling. *Phys. Rev. Lett.* **90,** 4 (2003).

33. Yang, S., Khare, K. & Lin, P. C. Harnessing surface wrinkle patterns in soft matter. *Adv. Funct. Mater.* **20,** 2550–2564 (2010).

34. Brakke, K. A. *Surface Evolver Documentation*. **2.7,** (2013).

35. Style, R. W. *et al.* Patterning droplets with durotaxis. *Proc. Natl. Acad. Sci.* **110,** 12541–12544 (2013).